\documentclass[conference]{IEEEtran}
\IEEEoverridecommandlockouts
\usepackage[utf8]{inputenc}
\usepackage{graphicx}
\usepackage{hyperref}
\usepackage[acronym]{glossaries}
\usepackage{multirow}
\usepackage{stackengine,scalerel}
\usepackage{accents}
\usepackage{algorithm}
\usepackage{algpseudocode}
\usepackage{gensymb}
\usepackage{cite}
\usepackage{amsmath,amssymb,amsfonts}
\usepackage{graphicx}
\usepackage{textcomp}
\usepackage{xcolor}
\usepackage{optidef}
\usepackage{accents} 
\usepackage{multirow}

\makenoidxglossaries
\glsdisablehyper   % Comment it to enable links for the acronyms

\newacronym{FD-ISAC}{FD-ISAC}{Full Duplex Integrated Sensing and Communication}
\newacronym[]{ISAC}{ISAC}{integrated sensing and sommunication}
\newacronym{ULA}{ULA}{uniform linear array}
\newacronym{UPA}{UPA}{uniform planar array}
\newacronym{MUSIC}{MUSIC}{MUltiple SIgnal Classification}
\newacronym{DL}{DL}{downlink}
\newacronym{WP2}{WP2}{Work Package 2}
\newacronym{RSU}{RSU}{road side unit}
\newacronym[]{TX}{TX}{transmitter}
\newacronym[]{RX}{RX}{receiver}
\newacronym[]{UE}{UE}{User Equipment}
\newacronym[]{AoD}{AoD}{angle of departure}
\newacronym[]{AoA}{AoA}{angle of arrival}
\newacronym[]{SINR}{SINR}{signal-to-interference plus noise ratio}
\newacronym[]{MPC}{MPC}{Multi path component}
\newacronym[]{MIMO}{MIMO}{multiple input multiple output}
\newacronym[]{mmWave}{mmWave}{millimeter-Wave}
\newacronym[]{BF}{BF}{beamforming}
\newacronym[]{CRB}{CRB}{Cramer Rao Bound}
\newacronym{BS}{BS}{base station}
\newacronym{UL}{UL}{uplink}
\newacronym{HBF}{HBF}{Hybrid analog and digital Beamforming} 
\newacronym{OFDM}{OFDM}{Orthogonal Frequency Division Multiplexing}
\newacronym{BB}{BB}{baseband}
\newacronym{LOS}{LOS}{line of sight}
\newacronym{NLOS}{NLOS}{Non line of sight}
\newacronym{SI}{SI}{self-interference}
\newacronym{DoA}{DoA}{direction of arrival}
\newacronym{DoD}{DoD}{direction of departure}
\newacronym{SNR}{SNR}{signal to noise ratio}
\newacronym{FOV}{FOV}{field of view}
\newacronym{ADC}{ADC}{Analog-to-Digital Converter}
\newacronym{PAPR}{PAPR}{Peak-to-Average-Power-Ratio}
\newacronym{DFT}{DFT}{discrete fourier transform}
\newacronym{NSP}{NSP}{null space projection}
\newacronym{MSS}{MSS}{maximum signal strength}
\newacronym{5G}{5G}{fifth generation}
\newacronym{DFRC}{DFRC}{dual-function radar-communication}
\newacronym{PSD}{PSD}{positive semi definite}

\def\BibTeX{{\rm B\kern-.05em{\sc i\kern-.025em b}\kern-.08em
    T\kern-.1667em\lower.7ex\hbox{E}\kern-.125emX}}

\newcommand{\g}{\mathbf{g}}

\newcommand{\x}{\mathbf{x}}
\newcommand{\y}{\mathbf{y}}

\newcommand{\W}{\mathbf{W}}

\begin{document}

\title{Multi-Target Two-way Integrated Sensing and Communications with Full Duplex MIMO Radios}
\author{\IEEEauthorblockN{Muhammad Talha\IEEEauthorrefmark{1}, Besma Smida\IEEEauthorrefmark{1}, Md Atiqul Islam\IEEEauthorrefmark{2}, and George C. Alexandropoulos\IEEEauthorrefmark{3} }
\IEEEauthorblockA{
{\IEEEauthorrefmark{1}Department of Electrical and Computer Engineering, University of Illinois at Chicago, USA}\\
{\IEEEauthorrefmark{2}Qualcomm Technologies, Inc., Santa Clara, CA, USA}\\
{\IEEEauthorrefmark{3}Department of Informatics and Telecommunications, National and Kapodistrian University of Athens, Greece}\\
emails: \{mtalha7, smida\}@uic.edu, mdatiqul@qti.qualcomm.com, alexandg@di.uoa.gr
}}

\maketitle

\begin{abstract}
In this paper, we propose a \gls{MIMO} Full-Duplex Integrated Sensing and Communication System consisting of multiple targets, a single downlink, and a single uplink user. We employed \gls{SINR} as the performance metric for radar, downlink, and uplink communication. We use a communication-centric approach in which communication waveform is used for both communication and sensing of the environment. We develop a sensing algorithm capable of estimating \gls{DoA}, range, and velocity of each target. We also propose a joint optimization framework for designing A/D transmit and receive beamformers to improve radar, downlink, and uplink \glspl{SINR} while minimizing \gls{SI} leakage. We also propose a \gls{NSP} based approach to improve the uplink rate. Our simulation results, considering \gls{OFDM} waveform, show accurate radar parameter estimation with improved downlink and uplink rate. 
\end{abstract}


\section{Introduction}
The ongoing discussions for the next generation of wireless networks pronounce the role of sensing, both as a means to effectively optimize spectral and energy efficiencies, as well as a core prerequisite characteristic of future immersive applications~\cite{Samsung}. In addition, sensing is envisioned to be offered in an integrated manner with next-generation communication services, giving birth to Integrated Sensing and Communications (ISAC) systems~\cite{mishra2019toward} that need to be supported by tailor-made physical-layer technologies~\cite{FD_survey_2023,RIS_ISAC_SPM}. Among those technologies belong the in-band Full-Duplex \gls{MIMO} systems~\cite{FD_MIMO_VTM2022}, which are capable of implementing simultaneous transmission and reception operations in the same frequency band. 

\textbf{Prior Works.} Leveraging an FD \gls{MIMO} system architecture with reduced complexity analog cancellation of the \gls{SI} signal\cite{alexandropoulos2017joint}, the authors in \cite{MultiuserComms-CE2020} jointly designed the \gls{BS} transmit and receive \gls{BF} matrices, as well as the settings for the multiple analog \gls{SI} cancellation taps and the digital Self-Interference (SI) canceller, to maximize the \gls{DL} rate simultaneously with the optimization of the \gls{UL} channel estimation process. FD MIMO radios have been also leveraged for efficient low-latency beam management in \gls{mmWave} systems. %In particular, \cite{Direction-Aided2020} presented a direction-assisted beam management framework, where the \gls{BS} was equipped with a large antenna array realizing \gls{DL} analog
%\gls{BF} and few digitally controlled reception antenna elements used for \gls{UL} estimation of the \gls{DoA} of the \gls{UL} signal from an intended user. 
A multi-beam \gls{mmWave} FD ISAC system was presented in~\cite{barneto2020beamforming}, according to which the \gls{TX} and \gls{RX} beamformers were optimized to have multiple beams for both communications and monostatic sensing. In \cite{liyanaarachchi2021joint}, the \gls{RX} spatial signal was further used to estimate the range and angle profiles corresponding to the multiple targets. An ISAC system, where a massive MIMO BS equipped with \gls{HBF} is communicating with multiple \gls{DL} users and simultaneously estimates via the same signaling waveforms the \gls{DoA} as well as the range of radar targets, has been presented in~\cite{Comms-Target_Tracking2022}. A more involved joint optimization framework for designing the analog and digital \gls{TX} and \gls{RX} beamformers, as well as the active analog and digital \gls{SI} cancellation units, was presented in \cite{ISAC2022} with the objective to maximize the achievable \gls{DL} rate and the accuracy performance of the \gls{DoA}, range, as well as the relative velocity estimation of radar targets. In \cite{BF_Design_FD_ISAC2022}, a \gls{BF} optimization framework to maximize radar sensing and the \gls{DL} users gain patterns while eradicating the \gls{SI} leakage power through a \gls{NSP} method, was presented. Moreover, in \cite{TXprecoding} different \gls{TX} precoding techniques have been discussed for \gls{DFRC} systems to ensure radar and communication guarantees. Very recently, in~\cite{FD_HMIMO_2023}, an FD MIMO architecture incorporating metasurface-based holographic MIMO transceivers~\cite{HMIMO_survey_all} was optimized for near-field ISAC in sub-THz frequencies. In~\cite{FD_RIS_ISAC_2023}, a reconfigurable intelligent surface was deployed in the near-field region of an FD MIMO node and was jointly optimized with the MIMO digital \gls{BF} to enable simultaneous communications and sensing of passive objects, while efficiently handling SI.

\textbf{Contributions.} In this paper, we consider an \gls{FD-ISAC} system consisting of multiple radar targets, single \gls{DL} and single \gls{UL} communication user. Unlike the previous work in~\cite{ISAC2022}, where a sub-optimal \gls{TX} precoding was presented because of bad performance at high \gls{TX} \gls{SNR} regime, we present an optimal approach using the Lagrangian duality method and derive the solution for the case of a single \gls{RX} RF chain. Moreover, to increase the \gls{UL} rate, we propose an \gls{NSP}-based approach to mitigate the radar interference from the intended \gls{UL} user signal. A similar problem was investigated in \cite{FD_ISAC_UL_DL_2023}, however, only digital \gls{BF} was considered at the \gls{BS} and user, which is rather intractable for large antenna arrays. The presented numerical results showcase that the proposed \gls{FD-ISAC} scheme performs sufficiently well in terms of radar as well as \gls{DL} and \gls{UL} \glspl{SINR}, while suppressing the \gls{SI} power to a desired threshold. 

    The remainder of the paper is organized as follows: Section \MakeUppercase{\romannumeral 2} describes the system model. Section \MakeUppercase{\romannumeral 3} presents the sensing parameters estimation procedure (\gls{DoA}, range and velocity of each target). In Section \MakeUppercase{\romannumeral 4} optimization framework is provided to optimize the beamformer to increase the corresponding \glspl{SINR}. In Section \MakeUppercase{\romannumeral 5} numerical results are provided and finally, Section \MakeUppercase{\romannumeral 6} concludes the paper.   

    \textit{Notations}: Matrices are denoted in uppercase bold letters (i.e., $\mathbf{H}$) while vectors in lowercase bold letters (i.e., $\mathbf{w}$). Sets of real and complex numbers are represented by $\mathbb{R}$ and $\mathbb{C}$ respectively. $||.||_\text{F}$ represents the Frobenius norm and $||.||_2$ represents the $\ell-2$ norm.  $[\mathbf{A}]_{(i,:)}$ represents the $i$-th row of matrix $\mathbf{A}$. 
    
\section{System Model}
 \begin{figure*}[!t]
     \centering
     \includegraphics[width=1.6\columnwidth]{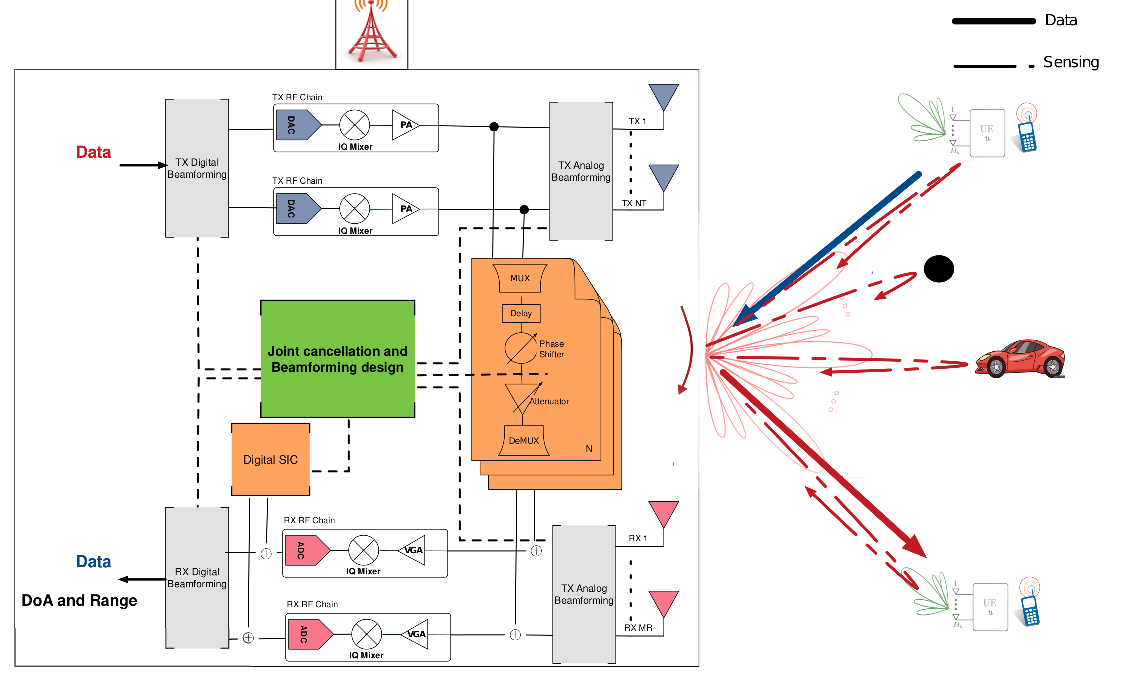}
     \caption{System Model for FD-ISAC -- This example set $K$=4, 1 UL user, 1 DL user and 2 passive targets.}
     \label{fig:sysmodel}
 \end{figure*}
{In the system model in the Fig. \ref{fig:sysmodel}, we consider \gls{FD-ISAC} communication system with a single \gls{BS} comprising of partially connected \gls{HBF} architecture at the \gls{TX} and \gls{RX} side and multiple targets, single \gls{DL} and single \gls{UL} user present in the vicinity of \gls{BS}. Specifically, \gls{BS} node $b$ consists of $N_b$ \gls{TX} and $M_b$ \gls{RX} antenna elements. The \gls{BS} is communicating with \gls{DL} user which consists of $M_u$ digital antenna elements and also with \gls{UL} user which consists of $N_u$ digital antenna elements. The \gls{DL} signal from the \gls{BS} also reflects back from the multiple passive radar targets present in the environment. Regarding the \gls{HBF} architecture at the \gls{BS} there are a total of $N_b^\text{RF}$ RF chains at the \gls{TX} and $M_b^\text{RF}$ RF chains at the \gls{RX}. Each RF chain is connected to $N_b^\text{A}$ analog antennas at the \gls{TX} and $M_b^\text{A}$ analog antennas at the \gls{RX}. 

  For the communication and sensing, \gls{OFDM} waveform is considered consisting of $P$ active subcarriers in each symbol and a total of $Q$ symbols. The frequency domain \gls{DL} and \gls{UL} symbol vectors at $p$-th sub-carrier and $q$-th symbol are represented as $\mathbf{s}_b^{p,q}$ and $\mathbf{s}_u^{p,q}$ respectively. At the \gls{BS}, \gls{BB} data vector is precoded using digital beamformer matrix  $\mathbf{V}_b^\text{BB} \in \mathbb{C}^{N_b^{RF} \times st}$ where $st$ represents data streams i.e., $st = \min \{N_b^{RF}, M_u\}$.  Following that, it is processed by analog beamformer $\mathbf{V}_b^\text{RF} \in \mathbb{C}^{N_b \times N_b^{RF}}$ containing a set of phase shifters as follows:
  
  \begin{equation}
  {\mathbf{V}}_b^{{\text{RF}}} = \left[ {\begin{array}{cccc} {{{\mathbf{v}}_1}}&{{{\mathbf{0}}_{N_b^{\text{A}} \times 1}}}& \cdots &{{{\mathbf{0}}_{N_b^{\text{A}} \times 1}}} \\ {{\mathbf{0}_{N_b^{\text{A}} \times 1}}}&{{{\mathbf{v}}_2}}& \cdots &{{{\mathbf{0}}_{N_b^{\text{A}} \times 1}}} \\ \vdots & \vdots & \ddots & \vdots \\ {{{\mathbf{0}}_{N_b^{\text{A}} \times 1}}}&{{{\mathbf{0}}_{N_b^{\text{A}} \times 1}}}& \cdots &{{{\mathbf{v}}_{N_b^{({\text{RF}})}}}} \end{array}} \right]. \label{eq:analogdesign}
  \end{equation}
The elements of each $\mathbf{v}_i$ are assumed to have a constant modulus, i.e., $|[\mathbf{v}_i]_n|^2 = 1/N_b^\text{A} \,\, \forall n = 1,2,\cdots,N_b^\text{A}$. It is also assumed that each vector in the analog beamformer belongs to some codebook, i.e., $\mathbf{v}_i \in \mathbb{F}_\text{TX} \forall i = 1,2, \cdots, N_b^\text{RF}$. The RF domain \gls{DL} signal at the \gls{BS} can be written as $\mathbf{x}_b^{p,q} = \mathbf{V}_b^\text{RF} \mathbf{V}_b^\text{BB} \mathbf{s}_b^{p,q},$ with power constraint: 
\begin{equation*}
    \mathbb{E}\left\{||\mathbf{V}_b^\text{RF} \mathbf{V}_b^\text{BB} \mathbf{s}_b^{p,q}||^2\right\} \leq \text{P}_b.
\end{equation*}
Similarly at the \gls{UL} user, after digital precoding, the transmitted vector can be represented as $ \mathbf{x}_u^{p,q} = \mathbf{V}_u^\text{BB}\mathbf{s}_u^{p,q}$, with power constraint:
\begin{equation*}
     \mathbb{E}\left\{||\mathbf{V}_u^\text{BB} \mathbf{s}_u^{p,q}||^2\right\} \leq \text{P}_u.
\end{equation*}
}
\subsection{Channel Models}{
    For \gls{DL} communication, it is assumed that there are $L$ scatterers present in the environment, which help to realize different \gls{LOS} multipaths in \gls{DL} channel. As \gls{mmWave} system is assumed, \gls{NLOS} paths are neglected. Hence, the \gls{DL} channel can be given as
    \begin{equation}
        \mathbf{H}_\text{DL} = \sum_{l=1}^L\alpha_l \mathbf{a}_{M_u}(\theta_l)\mathbf{a}_{N_b}^\text{H}(\theta_l),
    \end{equation}
    where $\alpha_l \in \mathbb{C}$ and $\theta_l \in [-90\degree , 90\degree]$ is the channel coefficient and angular direction of $l$-th path respectively. Moreover, $\mathbf{a}(\theta)$ is the \gls{ULA} response vector and is given by:
    \begin{equation*}
        \mathbf{a}_{N_b}(\theta) = \left[1,e^{j\frac{2\pi}{\lambda}d\sin(\theta)}, \cdots, e^{j\frac{2\pi}{\lambda}(N_b-1)d\sin(\theta)}\right]^\text{T},
    \end{equation*}
    where $\lambda$ is the signal wavelength and $d$ is the inter-element spacing of the antenna array. 
    For \gls{UL} communication, only one direct \gls{LOS} path is assumed. i.e., 
    \begin{equation}
        \mathbf{H}_\text{UL} = \beta \mathbf{a}_{M_b}(\phi)\mathbf{a}_{N_u}^\text{H}(\phi),
        \label{Channel:UL}
    \end{equation}
    where $\beta$ is the \gls{UL} channel coefficient and $\phi$ is the angular direction of user. 
    
    For sensing operation, it is assumed that there are $M$ passive radar targets, $L$ \gls{DL} scatterers, and a single LOS reflection from \gls{UL} user is considered. Hence total $K$ sensing targets are assumed i.e., $K = M + L +1$. The purpose of the sensing operation is to estimate the angular direction (Direction of Arrival (DoA)), range, and velocity of each radar target. The range and velocity of each target $k$ correspond to time delay $\tau_k$ and Doppler frequency $f_{D,k}$ respectively, such that $\tau_k = \frac{2 d_k}{c}$ and $f_{D,k} = 2 v_k f_c/c $, where $c$ is the speed of light, $v_k$ is the velocity of $k$-th target and $f_C$ is the carrier frequency. Hence the radar channel at $p$-th subcarrier and $q$-th OFDM symbol can be given as
    \begin{multline}
        \mathbf{H}_\text{Rad}^{p,q} = \sum_{k=1}^{K}\alpha_k e^{j2\pi \left(qT_sf_{D,k} - p\tau_k\Delta f\right)}\mathbf{a}_{M_b}(\theta_k)\mathbf{a}_{N_b}^\text{H}(\theta_k),
    \end{multline}
    where $\alpha_k$ is the reflection coefficient of the $k$-th radar target, $\Delta f$ is the subcarrier spacing and $T_s = \frac{1}{\Delta f} + T_{cp}$ is the total symbol duration (including the cyclic prefix) and $\theta_k$ is the angular direction of the $k$-th target. The two-way propagation delay due to the range of the target translates to a phase shift across each subcarrier and the Doppler shift due to the moving velocity translates to a phase shift across each OFDM symbol. Moreover, in order to be consistent with \gls{UL} channel, it is assumed that $\theta_K = \phi$. 
	}
 \subsection{Received Signal Models}
 {
 The \gls{DL} signal received by the user at the $p$-th subcarrier and $q$-th OFDM symbol can be written as
 \begin{equation}
     \mathbf{y}_u^{p,q} = \mathbf{W}_u^\text{H} \left(\mathbf{H}_\text{DL}\mathbf{x}_b^{p,q} + \mathbf{z}_u^{p,q}\right),
 \end{equation}
 where $\mathbf{W}_u$ is the digital \gls{RX} beamformer and $\mathbf{z}_u^{p,q} \sim \mathcal{CN}(\bold{0},\sigma_u^2\mathbf{I}_{M_u})$ is the AWGN noise at the user. 
    Similarly, the received signal at the \gls{RX} of \gls{BS} can be written as
    \begin{equation}
        \mathbf{y}_b^{p,q} = \underbrace{\mathbf{H}_\text{Rad}^{p,q}\mathbf{x}_b^{p,q}}_{\text{Radar Signal}} + \underbrace{\mathbf{H}_{b,b}\mathbf{x}_b^{p,q}}_{\text{\gls{SI} signal}} + \underbrace{\mathbf{H}_\text{UL}\mathbf{x}_u^{p,q}}_{\text{\gls{UL} signal}} + \mathbf{z}_b^{p,q},
    \end{equation}
    where $\mathbf{z}_b^{p,q} \sim \mathcal{CN}(\bold{0},\sigma_b^2\mathbf{I}_{M_b})$ is the additive noise at the \gls{BS} and $\mathbf{H}_{b,b} \in \mathbb{C}^{M_b\times N_b}$ is the \gls{SI} channel path between \gls{BS} \gls{TX} and \gls{RX}. The received signal is then processed with RF combiner matrix at the \gls{RX}, which is designed similarly as \ref{eq:analogdesign}. After RF combiner and A/D \gls{SI} cancellation, the received signal is expressed as
    \begin{multline}\label{eq: rx_sig_RF}
        \widetilde{\mathbf{y}}_b^{p,q} =\left(\mathbf{W}_b^{RF}\right)^\text{H}\left( \mathbf{H}_\text{Rad}^{p,q}\mathbf{x}_b^{p,q} + \mathbf{H}_\text{UL}\mathbf{x}_u^{p,q}\right) +\\ \left(\widetilde{\mathbf{H}}_{b,b}+\mathbf{C}_b + \mathbf{D}_b\right)\mathbf{V}_b^\text{BB}\mathbf{s}_b^{p,q} +\left(\mathbf{W}_b^\text{RF}\right)^\text{H}\mathbf{z}_b^{p,q}    ,
        \end{multline}
    where $\widetilde{\mathbf{H}}_{b,b} = \left(\mathbf{W}_b^\text{RF}\right)^\text{H}\mathbf{H}_{b,b}\mathbf{V}_b^\text{RF}$ and $\mathbf{C}_b$ and $\mathbf{D}_b$ are analog and digital cancellers and are designed by following the procedure presented in \cite{alexandropoulos2017joint}.  The expressions of $\mathbf{C}_b$ and $\mathbf{D}_b$ are derived based on the SI channel estimation $\widehat{\widetilde{\mathbf{H}}}_{b,b}$ as $\mathbf{C}_b = \Bigl[-\bigl[\widehat{\widetilde{\mathbf{H}}}_{b,b}\bigr]_{(:,1:\frac{N}{M_b^\text{RF}})} \, \boldsymbol{0}_{(:,N_b^{RF} - \frac{N}{M_b^{\text{RF}}}:N_b^\text{RF})}\Bigr]$ and $\mathbf{D}_b = -\bigl(\widehat{\widetilde{\mathbf{H}}}_{b,b} + \mathbf{C}_b\bigr)$, where $N$ is the number of analog canceller taps. \\
% \textcolor{blue}{[Talha please correct the above equation to include A/D SI cancellation]}
In order to extract the \gls{UL} signal at the \gls{BS} node $b$, a digital precoder is also applied at the \gls{RX} of the node $b$. Hence the received signal after the digital combiner can also be written as
    \begin{multline}
        \widetilde{\mathbf{s}}_b^{p,q} =\left(\mathbf{W}_b^\text{BB}\right)^\text{H}\left(\mathbf{W}_b^{RF}\right)^\text{H}\bigl(\mathbf{H}_\text{Rad}^{p,q}\mathbf{x}_b^{p,q} + \mathbf{H}_\text{UL}\mathbf{x}_u^{p,q}\bigr) + \left(\mathbf{W}_b^\text{BB}\right)^\text{H} \\ \left(\widetilde{\mathbf{H}}_{b,b} + \mathbf{C}_b + \mathbf{D}_b\right)\mathbf{V}_b^\text{BB}\mathbf{s}_b^{p,q} + \left(\mathbf{W}_b^\text{BB}\right)^\text{H}\left(\mathbf{W}_b^\text{RF}\right)^\text{H}\mathbf{z}_b^{p,q}.    \nonumber
    \end{multline}
 
 }

% \textcolor{blue}{[Talha please add another equation providing RX signal with digital RX beamforming]}

%%%%%%% Section III: Sensing Parameter Estimation -- Atiq (Update in progress)%%%%%%%%%%%

\section{Sensing Parameter Estimation}
In this section, we present the parameter estimation of the sensing targets utilizing the received signal at the BS node $b$. We estimate the DoA, range, and velocity of the targets using both reflected and uplink signals at the BS RX.\\

First, we estimate the \glspl{DoA} of targets including the direction of the DL and UL user using \gls{MUSIC} \cite{music} parameter estimation algorithm. After receiving the $PQ$ snapshots of the reflected signal, we calculate the sample covariance matrix of \eqref{eq: rx_sig_RF}. As the covariance matrix is positive semi-definite, after eigenvalue decomposition, all the eigenvalues will be real and non-negative. Now assuming $K < M_b^\text{RF}$, the subspace corresponding to the first $K$ eigenvalues will be signal space while the remaining subspace will correspond to noise subspace. In order to find the \glspl{DoA}, $\widehat{\theta_k},\forall k$, we calculate the vectors orthogonal to noise subspace by sweeping through all the angles in angular range.

Second, we estimate the range and velocity of the radar targets, including DL and UL users, by finding out the delay and Doppler shift associated with the $K$ estimated DoAs. We formulate a reference signal in the $k$th target direction as
\begin{equation}
    \begin{split}
        \g_{b,k}^{p,q} = \mathbf{a}_{M_b}(\widehat{\theta}_k)\mathbf{a}_{N_b}^{\rm H}(\widehat{\theta}_k)\x_{b}^{p,q}.
    \end{split}
\end{equation}
The reference signal $\g_{b,k}^{p,q}$ in the direction of $\widehat{\theta}_k$ and the RX signal $\widetilde{\y}_{p,q}$ in \eqref{eq: rx_sig_RF}  are utilized to derive quotient that includes the effect of delay and Doppler shift as
\begin{equation}
    \begin{split}
        z_{b,k}^{p,q} = \frac{1}{M_{b}}\sum\limits_{i=1}^{M_{b}}[\W_{b}^{\rm RF}\widetilde{\y}_{b}^{p,q}]_i/[\g_{b}^{p,q}]_i,\,\,\forall p,q.
    \end{split}
\end{equation}
Now, we estimate the quantized delay and Doppler shift associated with $k$th radar target as 
% \begin{equation}
%     \begin{split}
%         A(n,m) = \sum\limits_{p=0}^{P-1}\left(\sum\limits_{q=0}^{Q-1}
%             z_b^{p,q} e^{-j2\pi\frac{qm}{Q}}\right)e^{j2\pi\frac{pn}{P}},
%     \end{split}
% \end{equation}
\begin{equation}
    \begin{split}
        ({n}^*,{m}^*)\! = \!\underset{n,m}{\text{arg max}}\Big |\!\sum\limits_{p=0}^{P-1}\!\Big(\!\!\sum\limits_{q=0}^{Q-1}
            z_b^{p,q} e^{-j2\pi\frac{qm}{Q}}\Big)e^{j2\pi\frac{pn}{P}} \Big|^2
    \end{split}
\end{equation}
where $n = 0,\cdots,P-1$ and $m=-Q/2,\cdots,Q/2-1$. Finally, the estimated delay and Doppler frequency are expressed as $\widehat{\tau}_k = \frac{{n}^*}{P\Delta f}$ and $\widehat{f}_{D,k} = \frac{{m}^*}{QT_s}$, respectively.

%%%%%%%%%%%%%%%%%%%%%%%%%%%%%%%%%%%%%%%%%%%%%%%%%%%%%%%%%%%%%%%%%%%%%

\section{Proposed ISAC Optimization Framework}

In this section, we propose an optimization framework for designing A/D beamformers such that the DL rate, sensing performance and UL rate get maximized while constraining \gls{SI} power within a certain threshold. 

We consider a time division duplexing communication protocol, where the \glspl{DoA} estimated in one communication slot is utilized to derive the beamformers and \gls{SI} cancellation matrices in the successive slot. The performance metrics for \gls{DL} communication, \gls{UL} communication and sensing are corresponding \glspl{SINR}. The \gls{SINR} for sensing echo can be given as
\begin{equation}
				\gamma_\text{rad} = \frac{||\left(\mathbf{W}_b^\text{RF}\right)^H\widehat{\mathbf{H}}_\text{Rad}\mathbf{V}_b^\text{RF}\mathbf{V}_b^\text{BB}||_\text{F}^2}{||\left(	\widehat{\widetilde{\mathbf{H}}}_{b,b} + \mathbf{C}_b + \mathbf{D}_b \right)\mathbf{V}_b^\text{BB}||_\text{F}^2 + ||\mathbf{W}_b^\text{RF}||^2\sigma_b^2 }, \label{RadarSinr}
\end{equation}
where $ \widehat{\mathbf{H}}_\text{Rad} = \sum_{i-1}^{K-1}\mathbf{a}_{M_b}(\widehat{\theta}_i)\mathbf{a}_{N_b}^\text{H}(\widehat{\theta}_i) + \mathbf{a}_{M_b}(\widehat{\phi})\mathbf{a}_{N_b}^\text{H}(\widehat{\phi})$, $\widehat{\widetilde{\mathbf{H}}}_{b,b} = (\mathbf{W}_b^\text{RF})^\text{H}\mathbf{H}_{b,b}\mathbf{V}_b^\text{RF}$ and $\mathbf{C}_b$. It is to be noted that the \gls{UL} signal is not acting as the interference for the radar \gls{SINR}, because we consider \gls{UL} user as the active target present in the environment. Similarly the \gls{DL} \gls{SNR} is given by:
\begin{equation}
    \gamma_\text{DL} = \frac{||\mathbf{W}_{u}^H\widehat{\mathbf{H}}_\text{DL}\mathbf{V}_b^\text{RF}\mathbf{V}_b^\text{BB}||_\text{F}^2}{||\mathbf{W}_u||^2\sigma_{u}^2},
\end{equation}
where $\widehat{\mathbf{H}}_\text{DL} = \sum_{i=1}^L\mathbf{a}_{M_u}(\widehat{\theta}_i)\mathbf{a}_{N_b}^\text{H}(\widehat{\theta}_i)$.
Moreover, for \gls{UL} \gls{SINR}, a digital combiner is used after RF combiner in order to extract the transmitted signal by the \gls{UL} user. Hence the \gls{UL} \gls{SINR} is given by:
\begin{equation}
    \gamma_{\text{UL}} = \frac{||\left(\mathbf{W}_b^\text{BB}\right)^\text{H}\left(\mathbf{W}_b^\text{RF}\right)^\text{H}\widehat{\mathbf{H}}_\text{UL}\mathbf{V}_{u}^\text{BB}||_\text{2}^2}{\Sigma_\text{UL}} \label{eq:SINRUL}
\end{equation}
where $\Sigma_\text{UL} = ||\left(\mathbf{W}_b^\text{BB}\right)^\text{H}\left(\mathbf{W}_b^\text{RF}\right)^\text{H}\widehat{\mathbf{H}}_\text{Rad}\mathbf{V}_b^\text{RF}\mathbf{V}_b^\text{BB}||_\text{F}^2 + ||\left(\mathbf{W}_b^\text{BB}\right)^\text{H}\left(	\widehat{\widetilde{\mathbf{H}}}_{b,b} + \mathbf{C}_b + \mathbf{D}_b \right)\mathbf{V}_b^\text{BB}||_\text{F}^2 + \sigma_b^2 $ and $\widehat{\mathbf{H}}_\text{UL} = \mathbf{a}_{M_b}(\phi)\mathbf{a}_{N_u}^\text{H}(\phi)$. 

The optimization problem to maximize the sensing \gls{SINR}, \gls{DL} \gls{SNR} and \gls{UL} \gls{SINR} acn be written as
\begin{equation}
\begin{array}{c}
\mathcal{O}\mathcal{P}:\mathop {\max }\limits_{\begin{array}{l} {\mathbf{V}}_b^{{\text{RF}}},{\mathbf{V}}_b^{{\text{BB}}},{\mathbf{W}}_b^{{\text{RF}}},\mathbf{W}_b^\text{BB}, \\ {{\mathbf{C}}_b},{{\mathbf{D}}_b},\mathbf{V}_u^\text{BB},{{\mathbf{W}}_u} \end{array}} {{\gamma }_{{\text{Rad}}}} + {{\gamma }_{{\text{DL}}}} + \gamma_\text{UL} \\ {\text{s}}{\text{.t}}.{\left\| {{{\left[ {\left( {{{{\mathbf{\widehat {\widetilde H}}}}_{b,b}} + {{\mathbf{C}}_b}} \right){\mathbf{V}}_b^{{\text{BB}}}} \right]}_{(j,:)}}} \right\|^2} \leq {\lambda _b},\,\,\forall j = 1, \ldots ,M_b^{{\text{RF}}}, \\ {\mathbb{E}}\left\{ {{{\left\| {{\mathbf{V}}_b^{{\text{RF}}}{\mathbf{V}}_b^{{\text{BB}}}} \right\|}^2}} \right\} \leq {{\text{P}}_b} \\
\mathbb{E}\{||\mathbf{V}_u^\text{BB}||_2^2\} \leq \text{P}_{u}\\
{{\mathbf{w}}_j} \in {\mathbb{F}_{{\text{RX}}}}\forall j{\text{ and }}{{\mathbf{v}}_n} \in {\mathbb{F}_{{\text{TX}}}}\,\,\forall n = 1,2, \ldots ,N_b^{({\text{RF}})} 
\end{array}\label{optProblem:main} 
\end{equation}

Due to the coupling of optimization variables, the problem is non-convex and is challenging to solve. In this paper, an alternating optimization based solution is proposed. 

The analog \gls{TX} beamformer of \gls{BS} is optimized such that the radar channel gain (numerator of \eqref{RadarSinr}) gets maximized, which corresponds to a linear search through the codebook $\mathbb{F}_\text{TX}$. Similarly the optimization of \gls{BS} \gls{RX} beamformer is done such that radar channel return gets maximized and the \gls{SI} power leakage gets minimized. This is again a linear search through the codebook $\mathbb{F}_\text{RX}$. This is further elaborated in Algorithm \ref{alg:main}.

\subsection{\gls{TX} Digital Precoding}
{
    For designing digital \gls{TX} precoder, the goal is to maximize the \gls{DL} rate while minimizing \gls{SI} power at the \gls{RX} of node $b$. Hence following optimization problem is formalized:
    \begin{mini!}|s|[2]
    {\mathbf{V}_b^\text{BB}}
    {|| \widehat{\mathbf{H}}_\text{DL}^\text{eff}\mathbf{V}_b^\text{BB} - \mathbf{G}||_\text{F}^2 \label{optProblem2:Objective}} 
    {\label{optProblem2:Digital}}
    {}
    \addConstraint{||(\mathbf{V}_b^\text{BB})^\text{H}\mathbf{t}_r||_\text{2}^2 \,\,\leq \,\, \lambda_b \,\,\,\,\, \forall\,\,\, r = 1, \cdots , M_b^{RF}}{\label{optProblem2:con1}}
    \addConstraint{\Bigl|\Bigl|\left[\mathbf{V}_b^\text{RF}\mathbf{V}_{b}^\text{BB}\right]_{(:,c)}\Bigr|\Bigr|^2\,\,\leq \,\,  \text{P}_b\,\,\,\,\,\forall\,\,\, c = 1, \cdots , st}{\label{optProblem2:con2}}       
    \end {mini!}
    where $\widehat{\mathbf{H}}_\text{DL}^\text{eff} = \widehat{\mathbf{H}}_\text{DL}\mathbf{V}_b^\text{RF}$, $\mathbf{G} = \widehat{\mathbf{H}}_\text{DL}^\text{eff}\mathbf{V}\sqrt{\text{P}_b}$, where $\mathbf{V}$ is the right singular vectors of $\widehat{\mathbf{H}}_\text{DL}$ corresponding to largest singular values and $\mathbf{t}_r$ is the $r$-th row of the matrix $\widehat{\widetilde{\mathbf{H}}}_{b,b} + \mathbf{C}_b$ i.e., $\mathbf{t}_r = \left[\widehat{\widetilde{\mathbf{H}}}_{b,b} + \mathbf{C}_b\right]_{(r,:)}$. This is a convex optimization problem and for $M_b^\text{RF} = 1$ solution is proposed using the Lagrangian method. The Lagrangian for this problem (ignoring constraint \ref{optProblem2:con2}) can be written as
    \begin{equation}
        \mathcal{L}(\mathbf{V}_b^\text{BB},\zeta) = ||\widehat{\mathbf{H}}_\text{DL}^\text{eff}\mathbf{V}_b^\text{BB} - \mathbf{G}||_\text{F}^2 + \zeta(||(\mathbf{V}_b^\text{BB})^\text{H}\mathbf{t}_1||_2^2 -\lambda_b) \label{lagrange}
    \end{equation}
    where $\zeta$ is the Lagrange multiplier. By taking the derivative with respect to each element of $\mathbf{V}_b^\text{BB}$ yields:
    \begin{equation*}
        \frac{\partial}{\partial\mathbf{V}_b^\text{BB}}\mathcal{L}(\mathbf{V}_b^\text{BB},\zeta) = \left(\widehat{\mathbf{H}}_\text{DL}^\text{eff}\right)^\text{H}\widehat{\mathbf{H}}_\text{DL}^\text{eff}\mathbf{V}_b^\text{BB} - \left(\widehat{\mathbf{H}}_\text{DL}^\text{eff}\right)^\text{H}\mathbf{G} + \zeta\mathbf{t}_1\mathbf{t}_1^\text{H}\mathbf{V}_b^\text{BB}
    \end{equation*}
    The derivative of (\ref{lagrange}) with respect to lagrange multiplier $\zeta$ can be given as
    \begin{equation*}
         \frac{\partial}{\partial\zeta}\mathcal{L}(\mathbf{V}_b^\text{BB},\zeta) = \mathbf{t}_1^\text{H}\mathbf{V}_b^\text{BB}(\mathbf{V}_b^\text{BB})^\text{H}\mathbf{t}_1 - \lambda_b.
    \end{equation*}
    Equating both of these partial derivatives to zero will yield the optimal precoder and Lagrange multiplier $\accentset{\ast}{\zeta}$ i.e.,
    \begin{equation}
        \accentset{\ast}{\mathbf{V}}_b^\text{BB} = \left[\left(\widehat{\mathbf{H}}_\text{DL}^\text{eff}\right)^\text{H}\widehat{\mathbf{H}}_\text{DL}^\text{eff} + \accentset{\ast}{\zeta} \mathbf{t}_1\mathbf{t}_1^\text{H}\right]^{-1}\left(\widehat{\mathbf{H}}_\text{DL}^\text{eff}\right)^\text{H}\mathbf{G} \label{optprecoder1}
    \end{equation}
    \begin{equation*}
         \mathbf{t}_1^\text{H}\mathbf{V}_b^\text{BB}(\mathbf{V}_b^\text{BB})^\text{H}\mathbf{t}_1 = \lambda_b
    \end{equation*}
    Solving for $\zeta$ yields:
    \begin{equation}
         \accentset{\ast}{\zeta} = \frac{1}{{\mathbf{t}_1^H\left((\widehat{\mathbf{H}}_\text{DL}^\text{eff})^\text{H}\widehat{\mathbf{H}}_\text{DL}^\text{eff}\right)^{-1}\mathbf{t}_1}}\max\left( \frac{||\mathbf{t}_1||}{\sqrt{\lambda_b}}-1,0\right) \label{optlambda}
    \end{equation}
    Inserting the $\accentset{\ast}{\zeta}$ value in \eqref{optprecoder1} will yield optimal digital precoder. Here we assume that the inverse of $\left((\widehat{\mathbf{H}}_\text{DL}^\text{eff})^\text{H}\widehat{\mathbf{H}}_\text{DL}^\text{eff}\right)$ exists. While solving the problem \ref{optProblem2:Objective}, the constraint (\ref{optProblem2:con2}) is not considered. This constraint is forced by normalizing each vector of the precoder matrix by $\text{P}_b$ if it does not satisfy the constraint (\ref{optProblem2:con2}) i.e.,
    \begin{equation}
                \left[\mathbf{V}_b^\text{RF}\accentset{\ast}{\mathbf{V}}_b^\text{BB}\right]_{(:,c)} =
                \frac{\left[\mathbf{V}_b^\text{RF}\accentset{\ast}{\mathbf{V}}_b^\text{BB}\right]_{(:,c)} }{\Bigl|\Bigl|\left[\mathbf{V}_b^\text{RF}\accentset{\ast}{\mathbf{V}}_b^\text{BB}\right]_{(:,c)}\Bigr|\Bigr|}\sqrt{\text{P}_b}\,\,\, \forall c = 1, \cdots, st
    \end{equation}
    
}
\subsection{RX Digital combiner}
{
    After radar processing of the received signal, a digital combiner is applied for the extraction of \gls{UL} signal. As mentioned in \eqref{eq:SINRUL}, the radar signal acts as an interference to the \gls{UL} signal. Hence in order to maximize \gls{UL} \gls{SINR} \eqref{eq:SINRUL}, \gls{NSP} method is proposed to minimize the radar interference. In order to do that virtual radar channel is formulated without considering the \gls{UL} user as a target i.e., 
    \begin{equation}
        \widehat{\mathbf{H}}_\text{Rad,Int} = \sum_{i=1}^{K-1}\mathbf{a}_{M_b}(\widehat{\theta}_i)\mathbf{a}_{N_b}^\text{H}(\widehat{\theta}_i)
    \end{equation}

    The optimization problem for designing digital \gls{RX} precoder then can be formulated as
        \begin{maxi!}|s|[2]
    {\mathbf{W}_b^\text{BB}}
    {|| \left(\mathbf{W}_b^\text{BB}\right)^\text{H}\underbrace{\left(\mathbf{W}_b^\text{RF}\right)^\text{H}\widehat{\mathbf{H}}_\text{UL}}_{\widehat{\mathbf{H}}_\text{UL}^\text{eff}}||_\text{F}^2 \label{optProblem3:Objective}} 
    {\label{optProblem3:Digital}}
    {}
    \addConstraint{ \left(\mathbf{W}_b^\text{BB}\right)^\text{H}\underbrace{\left(\mathbf{W}_b^\text{RF}\right)^\text{H}\widehat{\mathbf{H}}_\text{Rad,Int}}_{\widehat{\mathbf{H}}_\text{Rad,Int}^\text{eff}} = \boldsymbol{0}}{\label{optProblem3:con1}}       
    \end {maxi!}
    
    % Then the problem (\ref{optProblem3:Digital}) can be re formulated as
    % \begin{mini!}|s|[2]
    % {\mathbf{W}_b^\text{BB}}
    % {|| \mathbf{W}_b^\text{BB} - \mathbf{X}||_\text{F}^2 \label{optProblem4:Objective}} 
    % {\label{optProblem4:DigitalRxCombiner}}
    % {}
    % \addConstraint{ \left(\widehat{\mathbf{H}}_\text{Rad,Int}^\text{eff}\right)^\text{H}\mathbf{W}_b^\text{BB} = \boldsymbol{0}}{\label{optProblem4:con1}}       
    % \end {mini!}
    
    In order to maximize the (\ref{optProblem3:Objective}), the $\mathbf{W}_b^\text{BB}$ can be chosen as the left singular vectors of matrix $\mathbf{H}_\text{UL}^\text{eff}$ corresponding to largest singular values. Hence the objective function \eqref{optProblem3:Objective} can be reformulated into the least squares function i.e., $\underset{\mathbf{W}_b^\text{BB}}{\min}\, || \mathbf{W}_b^\text{BB} - \mathbf{X}||_\text{F}^2$, with the same constraint \eqref{optProblem3:con1}, where $\mathbf{X}$ is the left singular vectors of matrix $\widehat{\mathbf{H}}_\text{UL}^\text{eff}$ corresponding to largest singular values. This reformulated problem is a well-studied problem and the solution is given as the \gls{NSP} of the matrix $\left(\widehat{\mathbf{H}}_\text{Rad,Int}^\text{eff}\right)^\text{H}$ \cite{NonLinearProgramming}. The optimal digital combiner can be given as
    \begin{equation}
        \accentset{\ast}{\mathbf{W}}_b^\text{BB} = \left(\mathbf{I} - \mathbf{A}^\text{H}\left(\mathbf{A}\mathbf{A}^\text{H}\right)^{-1}\mathbf{A}\right)\mathbf{X},
        \label{optprecoderrx}
    \end{equation}
where $\mathbf{A} = \left(\widehat{\mathbf{H}}_\text{Rad,Int}^\text{eff}\right)^\text{H} $.

    Each vector in the designed matrix will further be normalized to the unit norm. Our procedure for solving the optimization problem \ref{optProblem:main} has been summarized in Algorithm \ref{alg:main}. 
    \begin{algorithm}[!t]
        \caption{\gls{FD-ISAC} Optimization}
        \label{alg:main}
        \textbf{Input}: $\widehat{\mathbf{H}}_{b,b}, \widehat{\mathbf{H}}_\text{DL}, \widehat{\mathbf{H}}_\text{UL},\widehat{\mathbf{H}}_\text{Rad}, \text{P}_b, \text{P}_u$ and $\widehat{\theta}_k$  (Using \gls{MUSIC} \cite{music}) $\forall k$\\
        \textbf{Output}: $\mathbf{V}_b^\text{RF},\,\mathbf{V}_b^\text{BB},\,\mathbf{W}_b^\text{RF},\,\mathbf{W}_b^\text{BB},\,\mathbf{W}_u,\,\mathbf{V}_u^\text{BB},\,\mathbf{C}_b,\,\mathbf{D}_b$
        \begin{algorithmic}[1]
            \State Set $\mathbf{W}_u^\text{BB}$ as the $st$ left singular vectors of $\widehat{\mathbf{H}}_\text{DL}$ corresponding to largest singular values.
            \State Set $ {\mathbf{V}}_b^{{\text{RF}}} = \mathop {\operatorname{argmax} }\limits_{{{\mathbf{v}}_j} \in {\mathbb{F}_{{\text{TX}}}}} {\left\| {{{{\mathbf{\widehat H}}}_{\text{Rad}}}{\mathbf{V}}_b^{{\text{RF}}}} \right\|^2} $.
            \State Set ${\mathbf{W}}_b^{{\text{RF}}} = \mathop {\operatorname{argmax} }\limits_{{{\mathbf{w}}_j} \in {\mathbb{F}_{{\text{TX}}}}} \frac{{{{\left\| {{{\left( {{\mathbf{W}}_b^{{\text{RF}}}} \right)}^{\text{H}}}{{{\mathbf{\widehat H}}}_{\text{R}}}{\mathbf{V}}_b^{{\text{RF}}}} \right\|}^2}}}{{{{\left\| {{{\left( {{\mathbf{W}}_b^{{\text{RF}}}} \right)}^{\text{H}}}{{{\mathbf{\widehat H}}}_{{\text{b}},{\text{b}}}}{\mathbf{V}}_b^{{\text{RF}}}} \right\|}^2}}}$.
            \State Construct $\widehat{\widetilde{\mathbf{H}}}_{b,b} = \left(\mathbf{W}_b^\text{RF}\right)^\text{H}\widehat{\mathbf{H}}_{b,b}\mathbf{V}_b^\text{RF}$ and $\widehat{\mathbf{H}}_\text{DL}^\text{eff} = \widehat{\mathbf{H}}_\text{DL}\mathbf{V}_b^\text{RF}$. 
            \State Set $\mathbf{C}_b$ and $\mathbf{D}_b$.
            \If{$M_b^{RF} = 1$}
                \State Set $\accentset{\ast}{\zeta}$ as defined in \eqref{optlambda}.
                \State Set ${\mathbf{V}}_b^\text{BB}$ as defined in \eqref{optprecoder1} after solving problem in \ref{optProblem2:Digital}.
            \Else
                \State Use numerical optimization techniques to solve problem in \ref{optProblem2:Digital}.
            \EndIf
            \State Set $\mathbf{V}_u^\text{BB}$ as the first right singular vector of $\widehat{\mathbf{H}}_\text{UL}$ corresponding to largest singular value. 
            \State Set $\mathbf{W}_b^\text{BB}$ as defined in \eqref{optprecoderrx} after solving problem \ref{optProblem3:Digital}.
        \end{algorithmic}
    \end{algorithm}

}
\section{Numerical Results}
{
    In this section, we present numerical results for our proposed FD-ISAC system. The simulation parameters have been provided in the table \ref{table:params}. In addition to that, a radio subframe of 1ms has been assumed for \gls{DL} communication. The \glspl{RX} have an effective dynamic range of 60 dB provided by 14-bit \gls{ADC} for a \gls{PAPR} of 10 dB. Therefore, the residual \gls{SI} power after analog cancellation at each RF chain has to be below -30 dBm to avoid signal saturation. Moreover, the \gls{SI} channel has been modeled as a Rician fading channel with a $\kappa$ value of 35 dB and path loss of 40 dB. For a \gls{BS} analog \gls{TX}/\gls{RX} beamformer, we consider a 5-bit beam codebook based on \gls{DFT} matrix. 
    \begin{table}
    
    \caption{Simulation Parameters}
    \centering
        \begin{tabular}{|c||c|c|}
         \hline
         \multicolumn{3}{|c|}{Simulation Parameters} \\
         \hline
         Parameter Name& Symbol &Parameter value\\
         \hline
          Carrier Frequency&   $f_c$  & 28 GHz\\
         \gls{TX} RF chains & $N_b^\text{RF}$&8\\
         \gls{RX} RF chains&   $M_b^\text{RF}$&8 \\
         \gls{TX} Antennas per RF chain &$N_b^A$ & 16\\
         \gls{RX} Antennas per RF chain    &$M_b^A$ & 16\\
         Active Subcarriers& P  & 792   \\
         OFDM Symbols& Q  & 14\\
         Symbol Duration & $T_s$& 8.92 $\micro s$\\
         RX Noise Floor & $\sigma_b^2, \sigma_u^2$&-90 dBm\\
         \gls{RX} RF saturation level & $\lambda_b$& -30 dBm\\
         \hline
        \end{tabular}
        \label{table:params}
    \end{table}
\subsection{Sensing Performance}
{
\begin{figure}[t!]
     \centering
     \includegraphics[width=9cm]{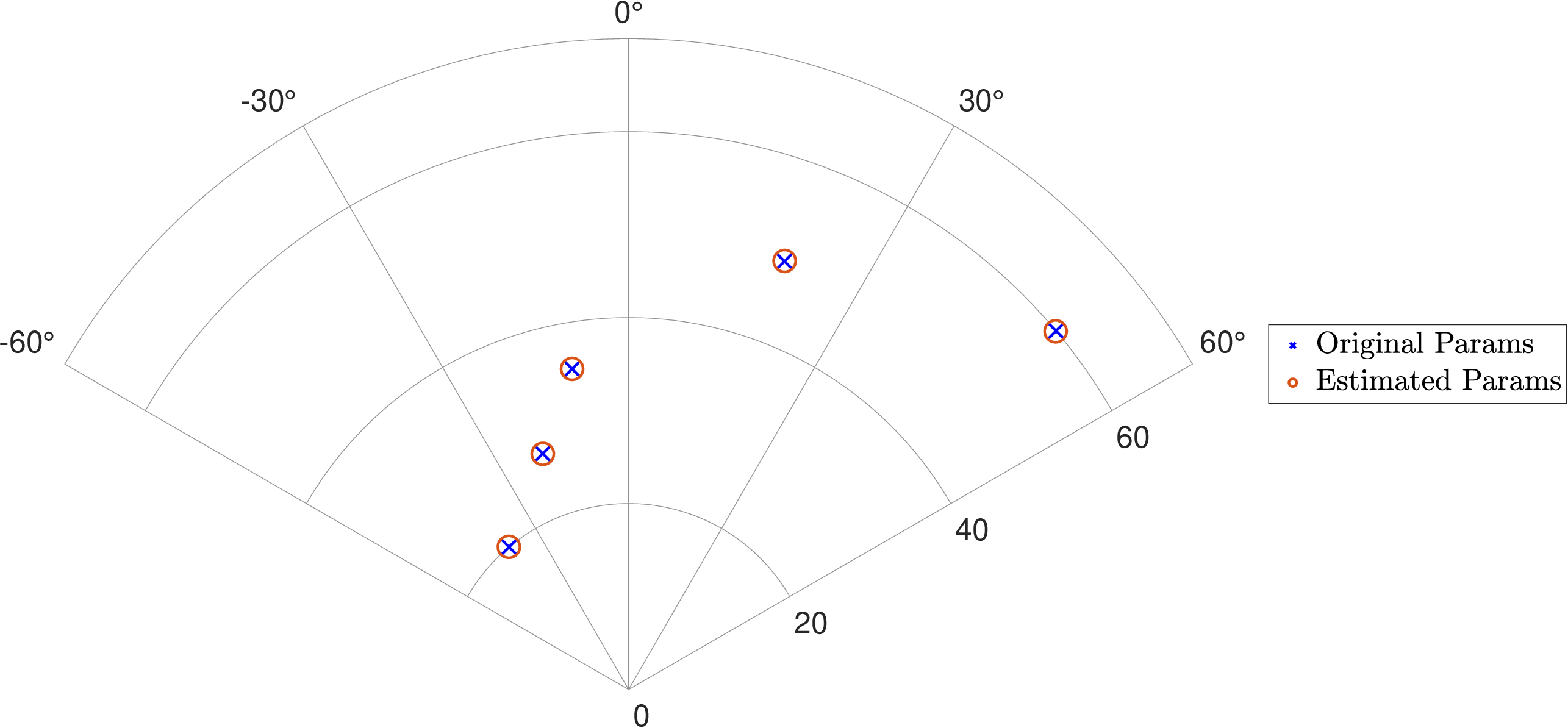}
     \caption{Range Angle Profile for \gls{TX} power of 30 dBm. The \gls{DL} link is associated with $L = 2$ scatteres at $\theta_1 = -30\degree$ and $\theta_2 = -20\degree$. The \gls{UL} user is present at the angle $\phi = -10\degree$.}
     \label{fig:RangeAngleMap}
 \end{figure}
 In Fig. \ref{fig:RangeAngleMap} the sensing performance of the proposed system is shown with a \gls{DL} transmit power of $30$ dBm and \gls{UL} transmit power of $10$ dBm. It can be seen that the estimated and original target parameters nearly overlap each other. Moreover, we also estimated the \gls{UL} user parameters (at $\phi = -10\degree$) without treating the \gls{UL} user as an interference to radar signal. Hence it showed that \gls{UL} user signal does not interfere with radar estimation as far as \gls{UL} user is treated as a potential target. The high angular resolution is due to MUSIC \gls{DoA} estimation, which is known to give highly accurate results as far $K < M_b^\text{RF}$. If $K > M_b^\text{RF}$ then one can use the conventional beam scanning approach to estimate \glspl{DoA}. 
\begin{figure}[t!]
     \centering
     \includegraphics[width=10.5cm]{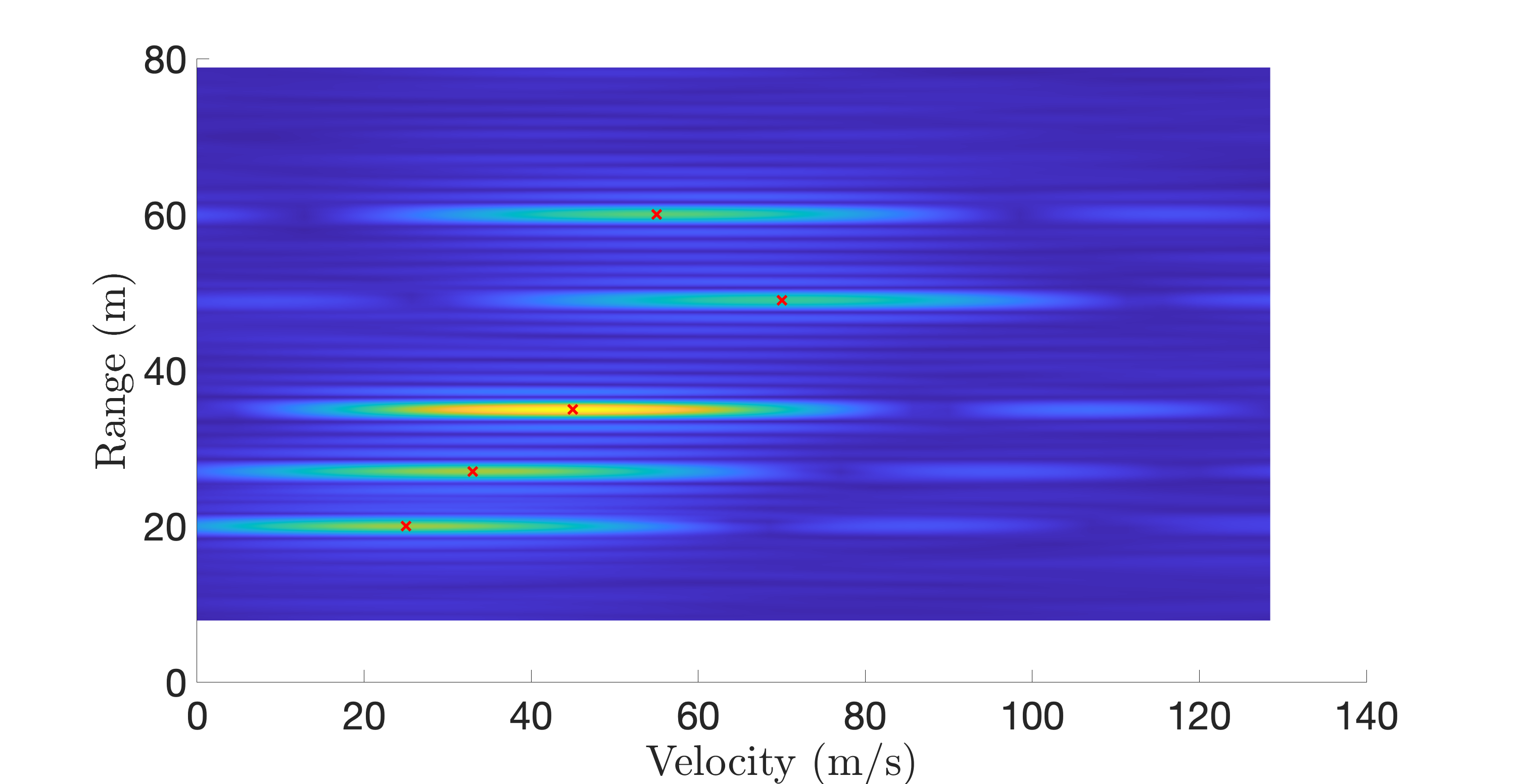}
     \caption{Range - Velocity Profile for \gls{TX} Power $\text{P}_b = 30$ dBm.}
     \label{fig:RangeVelocityMap}
 \end{figure}
 
    In Fig. \ref{fig:RangeVelocityMap} the range-velocity profile of the proposed system is shown. The map is obtained by implementing the range-velocity estimation algorithm individually for each target and then adding the maps after normalizing each one with the max value. It can be seen that the targets are easily recognizable across the range axis as compared to the velocity axis. This is because $Q << P$. Hence the number of samples across which the velocity is being measured is much less than the samples for range estimation. Applying the range and velocity estimation over the larger subset of \gls{OFDM} symbols will improve the velocity resolution. 
}
\subsection{\gls{DL} rate performance}
{
     \begin{figure}[t!]
     \centering
     \includegraphics[width=9.5cm]{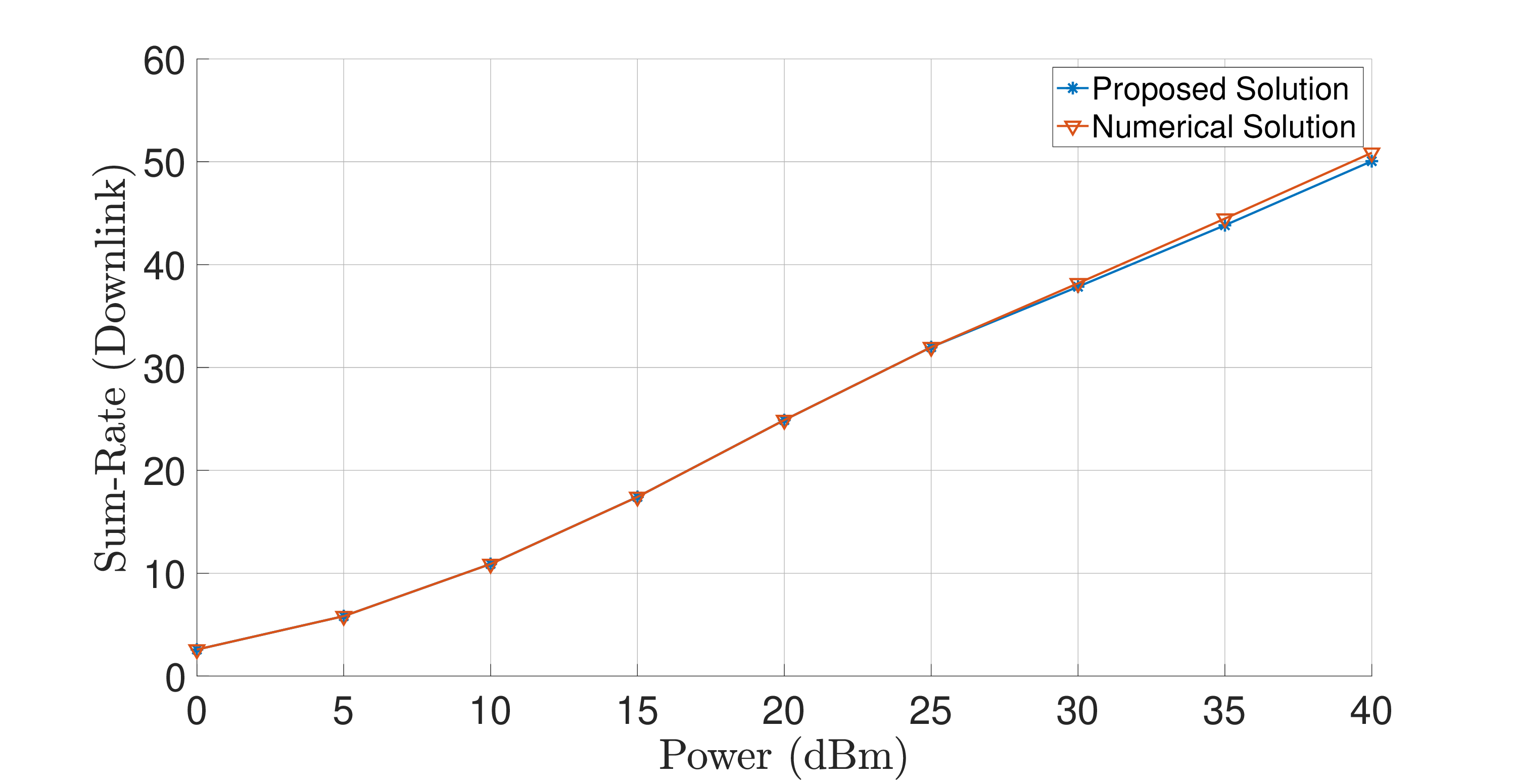}
     \caption{Effect pf proposed solution and numerical solution of \gls{TX} precoder design on Sum-Rate (bps/Hz) with $N_b^\text{RF} = 5,\, M_b^\text{RF} = 1,\, M_u = 5, \, L = 5 $. The rest of the system parameters are taken from table \ref{table:params}. The horizontal axis is the transmit power of \gls{BS} ($\text{P}_b$). }
     \label{fig:DlPrecoder}
 \end{figure}
      \begin{figure}[t!]
     \centering
     \includegraphics[width=9.5cm]{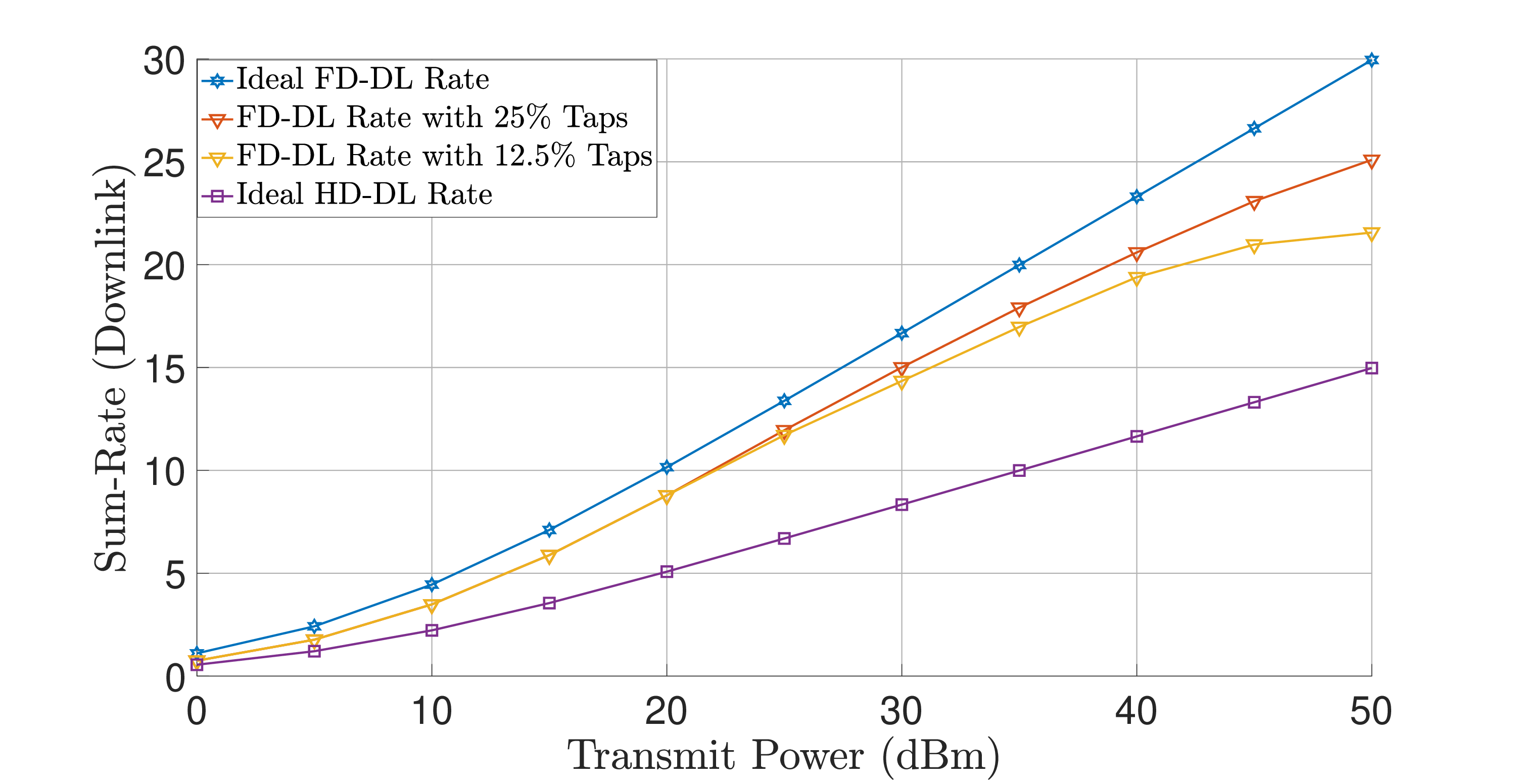}
     \caption{DL rate (bps/Hz) Performance using 128$\times$128 BS node and 4 antenna UE RX node}
     \label{fig:DLTaps}
 \end{figure}

  In Fig. \ref{fig:DlPrecoder}, the verification of the proposed \gls{DL} precoder is presented. \gls{DL} rate is taken as the performance metric for the \gls{DL} precoder design. It can be seen that the proposed solution is approximately matches with the numerical solution, which is computed through numerically solving the the convex optimization problem (\ref{optProblem2:Objective}) using CVX. Moreover, even when the \gls{SI} leakage is high (at high transmit power), our proposed algorithm manages to provide a considerably better \gls{DL} rate.  

  In Fig. \ref{fig:DLTaps}, DL rate performance is presented. For ideal rate calculation, a system having no \gls{SI} and radar channel is considered and all of beamformers are optimized in order to maximize the \gls{DL} rate. It can be seen that despite having \gls{SI} channel interference and radar targets in the environment, the proposed system is able to provide \gls{DL} rate which is very close to the ideal rate. Moreover, if we increase the analog \gls{SI} cancellation taps at the BS, DL rate and sensing performance improve as A/D beamformers are utilized to increase the rate and sensing performance. In the high transmit power regime, the \gls{SI} interference also scales up, and hence the system will utilize more resources to lessen the \gls{SI} signal and this will cause the \gls{DL} rate and sensing performance to deteriorate. 
}
}
\subsection{\gls{UL} rate performance}
{
      \begin{figure}[t!]
     \centering
     \includegraphics[width=9.5cm]{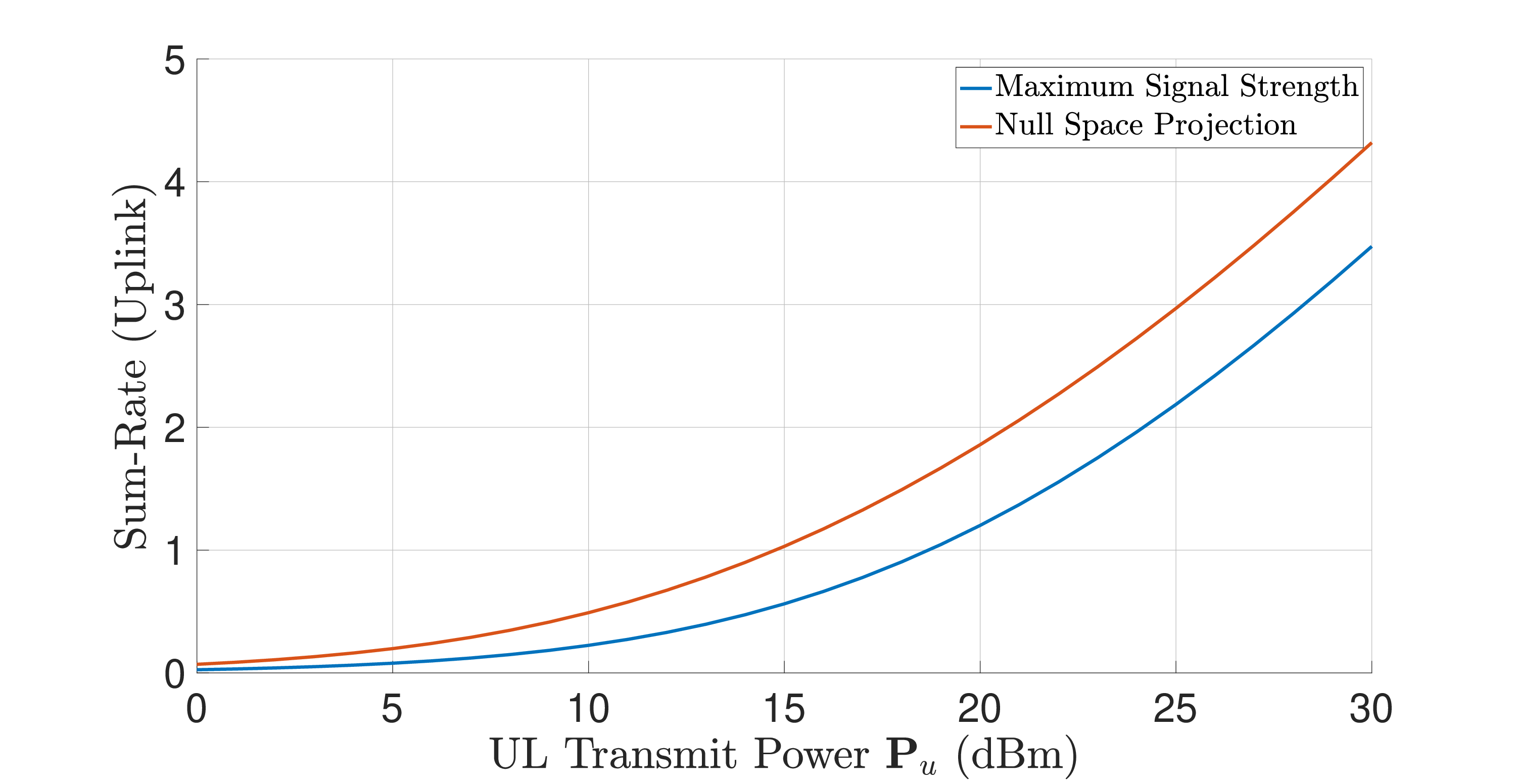}
     \caption{\gls{UL} rate (bps/Hz) performance with 4 antenna \gls{UE} \gls{TX} node and the \gls{BS} \gls{TX} power $\text{P}_b = 30 \text{ dBm}$.}
     \label{fig:ULRate}
 \end{figure}
Fig. \ref{fig:ULRate} depicts the \gls{UL} rate performance of the proposed system against \gls{UL} transmit power. The proposed \gls{NSP} method of receiver digital precoder has been compared with \gls{MSS} method where $\mathbf{W}_b^\text{BB}$ has been chosen as the left singular vectors of $\mathbf{H}_\text{UL}^\text{eff}$ without considering the constraint (\ref{optProblem3:con1}). It can be observed that the proposed method constantly provides a better \gls{UL} sum-rate than the \gls{MSS}. It is because, in \gls{MSS} as we are not forcing our precoder to belong to the null space of $\mathbf{H}_\text{rad,Int}^\text{eff}$, the radar signal act as a strong interference to the \gls{UL} signal which suppresses the \gls{UL} \gls{SINR} and ultimately the rate. 
}
\section{Conclusion}
{
   In this article, we presented \gls{FD-ISAC} communication system with multiple targets and \gls{UL} and \gls{DL} communication users. We proposed a complete framework for maximizing the \gls{DL} and \gls{UL} rate performance while also estimating the targets' radar parameters (\gls{DoA}, range and velocity). We optimize the A/D beamformers at both \gls{TX} and \gls{RX} of \gls{BS} and users to increase the corresponding \glspl{SINR} and to minimize the \gls{SI} leakage power at \gls{BS}. Our results demonstrated accurate radar parameter estimation and increased \gls{DL} and \gls{UL} rates. Moreover, our proposed algorithm manages to provide a high \gls{DL} rate even at high \gls{SI} leakage.

}

	\bibliography{reference}
	\bibliographystyle{ieeetr}
\vspace{12pt}

\end{document}